\documentclass[]{pasj00}
\def\vector#1{\mbox{\boldmath $#1$}}

\begin{document}
\SetRunningHead{Sato et al.}{Planetary Companions to Three Evolved
Intermediate-Mass Stars}
\Received{}
\Accepted{}

\title{Planetary Companions to Three Evolved Intermediate-Mass Stars:\\
HD 2952, HD 120084, and $\omega$ Serpentis}



%
 \author{
   Bun'ei \textsc{Sato},\altaffilmark{1}
   Masashi \textsc{Omiya},\altaffilmark{1}
   Hiroki \textsc{Harakawa},\altaffilmark{1}
   Yu-Juan \textsc{Liu}, \altaffilmark{2}
   Hideyuki \textsc{Izumiura},\altaffilmark{3,4}
   Eiji \textsc{Kambe},\altaffilmark{3}
   Yoichi \textsc{Takeda},\altaffilmark{4,5}
   Michitoshi \textsc{Yoshida},\altaffilmark{6}
   Yoichi \textsc{Itoh},\altaffilmark{7}
   Hiroyasu \textsc{Ando},\altaffilmark{4,5}
   Eiichiro \textsc{Kokubo}\altaffilmark{4,5}
   and
   Shigeru \textsc{Ida}\altaffilmark{1}
}
 \altaffiltext{1}{Department of Earth and Planetary Sciences, Tokyo Institute of Technology,
   2-12-1 Ookayama, Meguro-ku, Tokyo 152-8551, Japan}
 \email{satobn@geo.titech.ac.jp}
 \altaffiltext{2}{Key Laboratory of Optical Astronomy, National Astronomical Observatories,
   Chinese Academy of Sciences, Beijing 100012, China}
 \altaffiltext{3}{Okayama Astrophysical Observatory, National
   Astronomical Observatory of Japan, Kamogata,
   Okayama 719-0232, Japan}
 \altaffiltext{4}{The Graduate University for Advanced Studies,
   Shonan Village, Hayama, Kanagawa 240-0193, Japan}
 \altaffiltext{5}{National Astronomical Observatory of Japan, 2-21-1 Osawa,
   Mitaka, Tokyo 181-8588, Japan}
 \altaffiltext{6}{Hiroshima Astrophysical Science Center, Hiroshima University,
   Higashi-Hiroshima, Hiroshima 739-8526, Japan}
 \altaffiltext{7}{Nishi-Harima Astronomical Observatory, Center for Astronomy,
   University of Hyogo, 407-2, Nishigaichi, Sayo, Hyogo
   679-5313, Japan}
 

\maketitle

\begin{abstract}
We report the detections of planetary companions orbiting around three evolved
intermediate-mass stars from precise radial velocity measurements at
Okayama Astrophysical Observatory.
HD 2952 (K0III, 2.5$M_{\odot}$) and $\omega$ Ser (G8III, 2.2$M_{\odot}$) host a
relatively low mass planet with minimum mass of $m_2\sin i=1.6~M_{\rm J}$
and $1.7~M_{\rm J}$ in nearly circular orbits with period of $P=312$ and 277 d,
respectively.
HD 120084 (G7 III, 2.4$M_{\odot}$) hosts an eccentric planet with
$m_2\sin i=4.5~M_{\rm J}$ in an orbit with $P=2082$ d and eccentricity of
$e=0.66$.
The planet has one of the largest eccentricities among those ever discovered
around evolved intermediate-mass stars, almost all of which have eccentricity
smaller than 0.4.
We also show that radial velocity variations of stellar oscillations for G
giants can be averaged out below a level of a few m s$^{-1}$ at least in timescale
of a week by high cadence observations, which enables us to detect a
super-Earth and a Neptune-mass planet in short-period orbits even around
such giant stars.
\end{abstract}

\section{Introduction}\label{intro}
Planets and brown dwarfs discovered around intermediate-mass stars
(1.5--5$M_{\odot}$) have been growing in number for the last decade.
Hundreds of GK giants and subgiants, which are evolved counterparts
of intermediate-mass BA dwarfs, have been intensively surveyed
in radial velocity by utilizing their spectral features appropriate to
precise radial velocity measurements (e.g., \cite{frink:2002};
\cite{setiawan:2005}; \cite{hatzes:2005}; \cite{hatzes:2006};
\cite{lovis:2007}; \cite{niedzielski:2009b}; \cite{dollinger:2009};
\cite{demedeiros:2009}; \cite{sato:2010}; \cite{johnson:2011a};
\cite{wang:2012}; \cite{wittenmyer:2011}; \cite{omiya:2012};
\cite{lee:2012}; \cite{sato:2013}).
More than 50 substellar companions have been discovered around such
evolved intermediate-mass stars,
which are ranging from 0.6 to 40 $M_{\rm J}$ in minimum mass, from 0.08 to 6 AU
in semimajor axis, and from 0 to 0.68 in eccentricity. Planets less massive
than 2$M_{\rm J}$ were mostly found around subgiants because giants normally
show larger stellar jitter of 10--20 m s$^{-1}$ making detection of low mass planets
more difficult (e.g., \cite{sato:2005}), and all of the companions except
for only one hot-Jupiter around a subgiant (\cite{johnson:2010}) were
found in orbits beyond 0.6 AU.
Not only single planets but also multiple-planet systems have been
found, some of which are in mean-motion resonance
(\cite{johnson:2011b}; \cite{niedzielski:2009a}; \cite{niedzielski:2009b};
\cite{sato:2012}; \cite{sato:2013}).
Recently direct imaging succeeded in detecting
substellar companions in wide orbits around BA dwarfs, which are
complementary to radial velocity observations (e.g., \cite{marois:2008};
\cite{carson:2013}).
The orbital properties of these companions now serve as test cases for
general understanding of formation and evolution
of substellar companions around intermediate-mass stars (e.g., \cite{currie:2009}).

The Okayama Planet Search Program is one of the long-continued
planet search programs and has been regularly monitoring radial velocities
of about 300 intermediate-mass GK giants since 2001 at Okayama Astrophysical
Observatory (OAO) in Japan (\cite{sato:2005}). A total of 20 planets
and 6 brown dwarfs have been discovered so far at OAO including those
found in collaboration with Xinglong observatory in China, Bohyunsan Optical
Astronomy Observatory in Korea, Subaru 8.2m telescope in Hawaii,
and Australian Astronomical Observatory.
(e.g., \cite{sato:2012}; \cite{omiya:2012}; \cite{wang:2012}; \cite{sato:2010};
\cite{sato:2013}).

From the planet search program, here we report the discovery of three new
planetary companions around GK giants.
Rest of the paper is organized as follows. We describe the observations
in section \ref{obs} and the stellar properties are presented in
section \ref{stpara}.
Analyses of radial velocity, period search, orbital solution, stellar activity,
and line shape variation are described in section \ref{ana} and the results are
presented in section \ref{results}.
Section \ref{discussion} and \ref{summary} are devoted to discussion
and summary, respectively.

\section{Observation}\label{obs}
All of the observations were made with the 1.88 m reflector and the HIgh
Dispersion Echelle Spectrograph (HIDES; \cite{izumiura:1999}) at OAO.
In 2007 December, HIDES was upgraded from single CCD (2K$\times$4K)
to a mosaic of three CCDs, which enables us to simultaneously obtain
spectra with a wavelength range of 3750--7500${\rm \AA}$ using a red
cross-disperser. Furthermore, the high-efficiency fiber-link
system with an image slicer has been available for HIDES since 2010,
which makes overall throughput
more than twice than that with the conventional slit observations
(\cite{kambe:2013}).
We basically obtained the data presented in this paper using the
slit (hereafter HIDES-Slit) mode, but we also used the fiber-link
(hereafter HIDES-Fiber) mode for two of the three stars.

In the HIDES-Slit mode, the slit width was set to 200 $\mu$m
($0.76^{\prime\prime}$) giving a spectral resolution
($R=\lambda/\Delta\lambda$) of 67000 by about 3.3 pixels sampling.
In the HIDES-Fiber mode, the width of the sliced image is $1.05^{\prime\prime}$
corresponding to a spectral resolution of $R=$55000 by about 3.8
pixels sampling. Each observing mode uses its own iodine absorption cell
(I$_2$ cell; \cite{kambe:2002, kambe:2013}) for precise radial velocity
measurements, which provides a fiducial wavelength reference
in a wavelength range of 5000--5800${\rm \AA}$. Possible offset in
radial velocities between the two modes caused by using the different
I$_2$ cells is treated as a free parameter in orbital fitting
(see section \ref{orbit}).

The reduction of echelle data (i.e. bias subtraction, flat-fielding,
scattered-light subtraction, and spectrum extraction) is performed
using the IRAF\footnote{IRAF is distributed by the National
Optical Astronomy Observatories, which is operated by the
Association of Universities for Research in Astronomy, Inc. under
cooperative agreement with the National Science Foundation,
USA.} software package in the standard way.

\section{Stellar Properties}\label{stpara}
Atmospheric parameters (effective temperature $T_{\rm eff}$,
surface gravity $\log g$, micro-turbulent velocity $v_t$, Fe
abundance [Fe/H], and projected rotational velocity $v\sin i$)
of all the targets for Okayama Planet Search
Program were derived by \citet{takeda:2008}, based on the
measured equivalent widths of well-behaved Fe I and Fe II lines of
iodine-free stellar spectra.
Details of the procedure and resultant parameters are presented
in \citet{takeda:2002} and \citet{takeda:2008}.

\begin{figure}
  \begin{center}
    \FigureFile(85mm,80mm){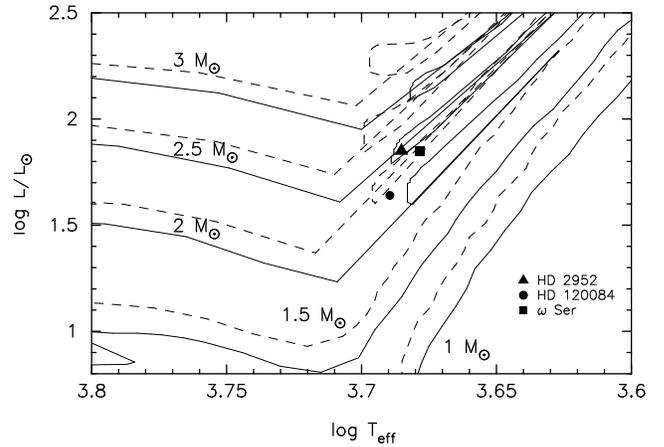}
  \end{center}
\caption{HR diagram of the planet-harboring stars presented in this paper.
Pairs of evolutionary tracks from Lejeune and Schaerer (2001)
for stars with $Z=0.02$ (solar metallicity; solid
lines) and $Z=0.008$ (dashed lines) of masses between 1 and 3
$M_{\odot}$ are also shown.}\label{fig-HRD}
\end{figure}

They also obtained the absolute magnitude $M_V$ of each star
from the apparent $V$-band magnitude and Hipparcos parallax
$\pi$ (\cite{esa:1997}) correcting interstellar extinction
$A_V$ from \citet{arenou:1992}'s table, and obtained the bolometric
correction $B.C.$ based on the \citet{kurucz:1993}'s theoretical
calculation.
The luminosity $L$ and mass $M$ of each star were derived
using these parameters and theoretical evolutionary tracks of
\citet{lejeune:2001}, and the stellar radius $R$ was derived
by the Stefan-Boltzmann relationship and the measured $L$ and $T_{\rm eff}$.
The properties of the three stars presented in this paper (HD 2952,
HD 120084, $\omega$ Ser) are summarized in table \ref{tbl-stars}, and
the stars are plotted on the HR diagram in figure \ref{fig-HRD}.

\begin{figure}
  \begin{center}
    \FigureFile(85mm,80mm){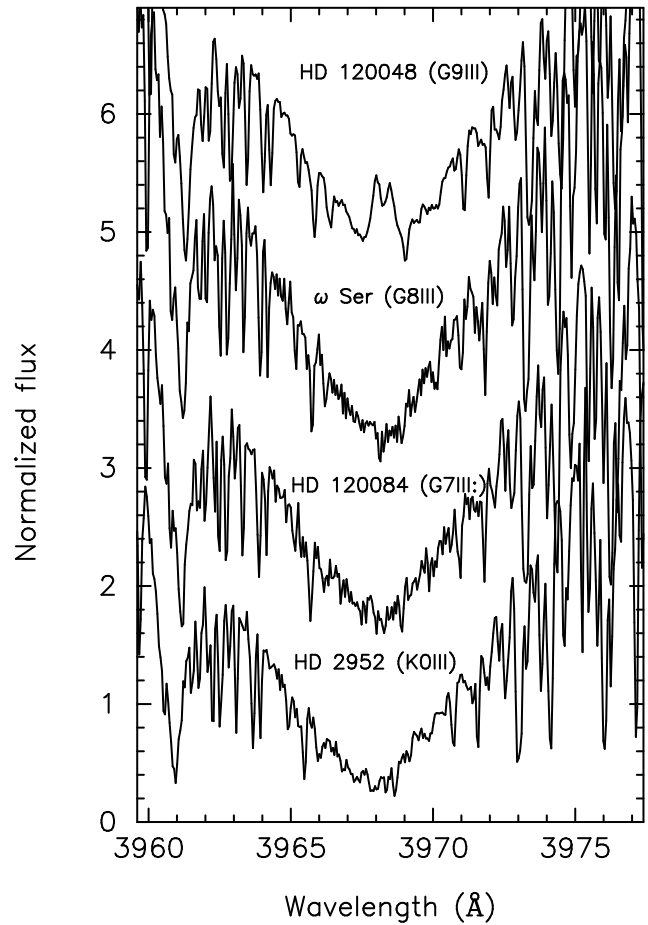}
  \end{center}
\caption{Spectra in the region of Ca H lines. All of the stars show no
significant core reversals in the lines compared to that in the
chromospheric active star HD 120048.
A vertical offset of about 0.8 is added to each spectrum.}\label{fig-CaH}
\end{figure}

The three stars are known to be stable in photometry to a level
of $\sigma_{\rm HIP}=0.005-0.007$ mag (\cite{esa:1997}), and
they are chromospherically inactive with no significant emission
in the core of Ca II HK lines as shown in figure \ref{fig-CaH}.
Since we can obtain spectra covering Ca II HK lines together
with radial velocity data after installing 3 CCDs in 2007,
we use the lines to check stellar chromospheric activity correlated
with radial velocity variations (see section \ref{activity}).

\section{Analysis}\label{ana}
\subsection{Radial Velocity}\label{rv ana}
Radial velocity analysis was carried out using the modeling
technique of an I$_2$-superposed stellar spectrum detailed
in \citet{sato:2002} and \citet{sato:2012}, which is based
on the method by \citet{butler:1996}.
In the technique, an I$_2$-superposed
stellar spectrum is modeled as a product of a high resolution
I$_2$ and a stellar template spectrum convolved with a
modeled instrumental profile (IP) of the spectrograph.
We here used a stellar template that was obtained by
deconvolving a pure stellar spectrum with the spectrograph
IP estimated from an I$_2$-superposed B-star or Flat spectrum.
We applied the stellar template thus obtained with HIDES-Slit mode
observations to radial velocity analysis for the data taken
with HIDES-Fiber mode as well as HIDES-Slit mode.
We took account of a velocity offset, $\Delta$RV$_{\rm f-s}$,
between the two observing modes as a free parameter when we
determine orbital parameters (see section \ref{orbit}).
Measurement error in radial velocity was estimated from an
ensemble of velocities from each of $\sim$300 spectral regions
(each $\sim$3${\rm \AA}$ long) in every exposure. 

\subsection{Period Search}\label{period search}
A Lomb-Scargle periodogram analysis (\cite{scargle:1982})
was performed to search for periodicity in radial velocity data,
and False Alarm Probability ($FAP$) was estimated
to assess the significance of the periodicity.
To estimate the $FAP$, we created 10$^5$ fake datasets,
in which the observed radial velocities were randomly redistributed,
keeping fixed the observation time, and applied the same periodogram
analysis to them. The fraction of fake datasets exhibiting
a periodogram power higher than the observed one was
adopted as a $FAP$ for the signal.

\subsection{Keplerian Orbital Solution by MCMC Method}\label{orbit}
Keplerian orbital model for the radial velocity data and uncertainties
for the parameters were derived using the Bayesian Markov chain Monte
Carlo (MCMC) method (e.g., \cite{ford:2005}; \cite{gregory:2005};
\cite{ford:2007}). Since the details of the method are presented in the
literatures, we here briefly describe the procedure and parameters for
the model adopted in this paper.

In Bayes' theorem, the target joint probability distribution, the posterior
probability distribution, for parameters \vector{\theta} of a certain model
$M$ based on observational data \vector{d} and prior background information
$I$ is given by
\begin{eqnarray}
p(\vector{\theta}|\vector{d}, I, M) &=& Cp(\vector{\theta}| I, M)\times p(\vector{d}|\vector{\theta}, I, M)
\end{eqnarray}
where $C$ is the normalization constant, $p(\vector{\theta}| I, M)$ is the
prior probability distribution of \vector{\theta}, and $p(\vector{d}|\vector{\theta}, I, M)$
is the likelihood, which is the probability that we would have obtained
the data \vector{d} given the parameters \vector{\theta}, model $M$,
and priors $I$.

The likelihood function is given by
\begin{eqnarray}
p(\vector{d}|\vector{\theta}, I, M) &=& A\exp\left[\sum_{j}\sum_{i=1}^{N_j}\frac{(v_{i,j}-y_{i,j})^2}{2\left(\sigma_{i,j}^2+s_j^2\right)}\right]\\
A &=& \prod_j\left(2\pi\right)^{-N_j/2}\prod_{i=1}^{N_j}\left(\sigma_{i,j}^2+s_j^2\right)^{-1/2}
\end{eqnarray}
where $N_j$ is the number of data points for the $j$th instrument,
$\sigma_{i,j}$ is the measurement uncertainties for each point,
$v_{i,j}$ and $y_{i,j}$ are observed and modeled radial velocities, respectively.
Extra Gaussian noise $s_j$ is incorporated for observations with $j$th
instrument, including intrinsic stellar jitter such as stellar oscillation and
unknown noise source, in addition to the measurement
uncertainties.

For a single planet, reflex motion in stellar radial velocity at time $t_i$
observed with $j$th instrument can be expressed as
\begin{eqnarray}
y_{i,j} (t_i) &=& V_j + K_1 [\sin (f_{i}+\omega) + e\sin\omega]
\end{eqnarray}
where $V_j$ is systematic velocity with reference to $j$th instrument,
$K_1$, $\omega$, $e$, and $P$ is velocity semiamplitude,
argument of periastron, eccentricity, and orbital period,
respectively. $f_{i}$ is true anomaly of the planet at $t_i$,
which is a function of $e$, $P$, and time of periastron passage $T_p$.
Instead of $T_p$, we here adopted $\chi$, fraction of an orbit of the
planet prior to the start of data taking, at which periastron occurred.

For the prior probability distribution, we follwed \citet{ford:2007}
adopting Jeffrey's prior for $P$, modified
Jeffrey's prior for $K_1$ and $s_j$, and uniform one for other parameters.
Then, joint prior for the model parameters, assuming independence,
is given by
\begin{eqnarray}
p(\vector{\theta}| I, M)= p(P|I,M)p(K_1|I,M)p(e|I,M)p(\omega|I,M)\nonumber\\
\times p(\chi|I,M)\prod_j p(V_j|I,M)p(s_j|I,M)\nonumber\\
=\frac{1}{P\ln(P_{max}/P_{min})}\frac{1}{(K_1+K_a)\ln[(K_a+K_{max})/K_a]}\nonumber\\
\times\frac{1}{2\pi}\prod_j\frac{1}{V_{j,max}-V_{j,min}}\frac{1}{(s_j+s_a)\ln[(s_a+s_{max})/s_a]}.
\end{eqnarray}
The parameter priors we adopted are summarized in table \ref{tbl-priors}.

The posterior probability distribution was obtained with the MCMC method,
which uses a stochastic process to generate a ``chain'' of points in
parameter space that approximates the desired probability distribution.
The chain is derived from an initial point by iterating a ``jump function'',
which was the addition of a Gaussian random number to each parameter value
in our case.
If the new set of parameters \vector{\theta^\prime} has a larger posterior
probability $p(\vector{\theta^\prime}|\vector{d}, I, M)$ than that for
the previous set of parameters $p(\vector{\theta}|\vector{d}, I, M)$,
the jump is executed and the new parameters are accepted; if not, the
jump is only executed with probability
$p(\vector{\theta^\prime}|\vector{d}, I, M)/p(\vector{\theta}|\vector{d}, I, M)$;
otherwise, the previous parameters are recorded. We set the relative sizes
of the Gaussian perturbations based on the 1$\sigma$ uncertainties
estimated on ahead with the bootstrap method, and required the
overall acceptance rate of $\sim$25\%, which is the total fraction of jumps
executed, as recommended by \citet{roberts:1997}.
We generated 5 independent chains, each of which started from random
initial values 5$\sigma$ away from the best-fit ones. Each chain had $10^6$--$10^7$ points,
the first 10\% or 500,000 of which were discarded to minimize the effect
of the initial condition. To check the convergence and the consistency between
the chains, we computed Gelman-Rubbin statistic (\cite{gelman:1992}) for each
parameter, which is a comparison between the interchain variance
and the intrachain variance, and confirmed that the results were
within a few percent of unity, a sign of good mixing and convergence.
We also calculated an estimate of the effective number of independent
draws following \citet{gelman:2003} and confirmed that the value was
larger than 1000. The resultant chains were merged to derive the final
joint posterior probability distribution function (PDF).
We derived the median value of the PDF for each parameter and set
the 1$\sigma$ uncertainty as the range between 15.87\% and 84.13\%
of the PDF.

\subsection{Stellar Activity}\label{activity}
Rotational modulation of spots or activity cycle of stars cause apparent
variations in stellar radial velocities, which can masquerade as a planetary
signal (e.g., \cite{queloz:2001}). Thus it is important to check the activity
level of stars together with radial velocity variations. For this purpose,
Ca II HK lines are widely used as activity indicators since the flux
of the line cores is a good tracer of chromospheric activity (e.g., \cite{duncan:1991};
\cite{noyes:1984}).
We here define Ca II H index $S_H$ as
\begin{eqnarray}
S_H &=& \frac{F_H}{F_B+F_R}
\end{eqnarray}
where $F_H$ is a total flux in a wavelength bin
0.66${\rm \AA}$ wide centered on the H line, $F_B$ and $F_R$
are those in bins 1.1 ${\rm \AA}$ wide centered on minus
and plus 1.2 ${\rm \AA}$ from the center of the H line, respectively.
By setting the reference wavelength bands, $B$ and $R$, close to
the base of the CaII H line core in this way, we tried to minimize
the error in $S_H$ caused by imperfect normalization of the continuum
level. We obtained the $S_H$ values only for spectra with S/N of
$F_H$ over $\sim$40 among those taken after 2007.

\subsection{Line Shape Variation}\label{line shape}
To investigate other possible causes of apparent radial velocity
variations rather than orbital motion, spectral line shape
analysis was performed using IP-deconvolved stellar templates.
Details of the analysis are described in \citet{sato:2007} and
\citet{sato:2002}, and here we briefly summarize the procedure.

At first, two IP-deconvolved templates were extracted from five
I$_2$-superposed stellar spectra at nearly the phases of maximum
and minimum in observed radial velocities using the technique by
\citet{sato:2002}.
Cross correlation profiles of the two templates were then derived
for about 100 spectral segments (4--5${\rm \AA}$
width each) that have no severely blended or broad lines in them.
Three bisector quantities were calculated for the cross
correlation profile for each segment:
the velocity span (BVS), the velocity difference between two flux
levels of the bisector; the velocity curvature (BVC), the
difference of the velocity span of the upper and lower half
of the bisector; and the velocity displacement (BVD), the
average of the bisector at three different flux levels.
Flux levels of 25\%, 50\%, and 75\% of the cross correlation
profile were used to calculate the above three quantities.
Both the BVS and the BVC being identical to zero and the
average BVD agreeing with the velocity difference between the two
templates ($\simeq 2K_1$) suggest that the cross correlation
profiles can be considered to be symmetric. Thus the observed
radial velocity variations are considered to be caused by
parallel shifts of the spectral lines rather than deformation
of them, which favors the orbital motion hypothesis.

\section{Results}\label{results}

\subsection{HD 2952 (HR 135, HIP 2611)}\label{HD2952}
We collected a total of 63 radial velocity data of HD 2952 between
2004 January and 2012 December using both of HIDES-Slit and HIDES-Fiber mode.
We obtained a signal-to-noise ratio S/N$=$100--260 pix$^{-1}$ at 5500${\rm \AA}$
with an exposure time 780--1800 sec using HIDES-Slit mode and
S/N$=$200--320 pix$^{-1}$ at 5500${\rm \AA}$ with an exposure time 450--900 sec
using HIDES-Fiber mode.

\begin{figure}
  \begin{center}
    \FigureFile(85mm,80mm){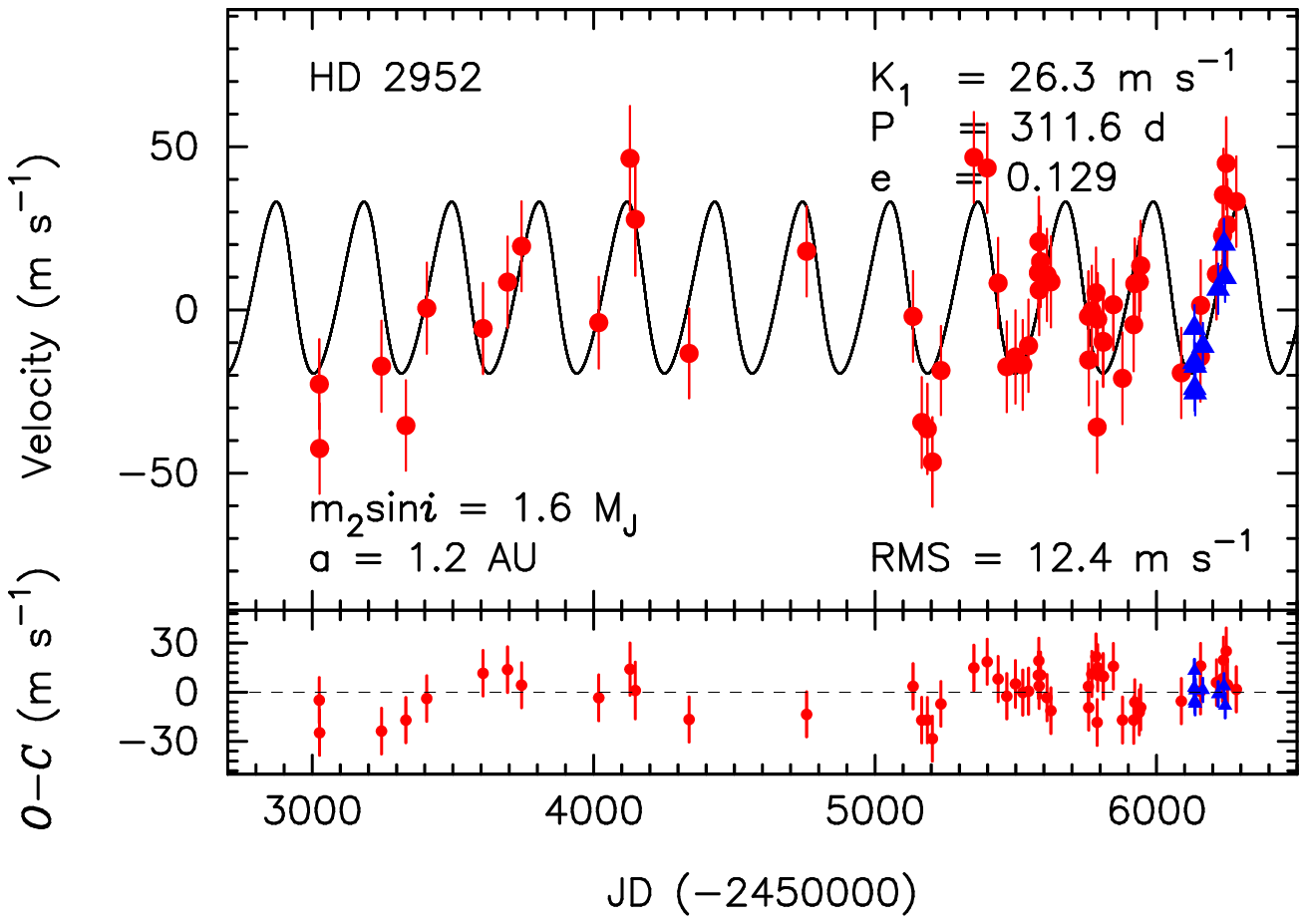}
  \end{center}
\caption{{\it Top}: Radial velocities of HD 2952 observed at OAO.
HIDES-Slit data are shown in filled red circles, and
HIDES-Fiber data are filled blue triangles.
The Keplerian orbit is shown by the solid line.
The error bar for each point includes the stellar jitter.
{\it Bottom}: Residuals to the orbital fit.
The rms to the fit is 12.4 m s$^{-1}$.}
\label{fig-HD2952}
\end{figure}

The observed radial velocities are shown in figure \ref{fig-HD2952}
and are listed in table \ref{tbl-HD2952} together with their estimated
uncertainties. Lomb-Scargle periodogram of the data exhibits a dominant
single peak at a period of 313 days with $FAP=1\times10^{-5}$.

The orbital parameters for single Keplerian model derived by MCMC method
are $P=311.6^{+1.7}_{-1.9}$ days, $K_1=26.3^{+3.8}_{-3.4}$ m s$^{-1}$,
and $e=0.129^{+0.099}_{-0.085}$, which are median and 68.3\% credible
regions for PDF.
The resulting Keplerian model is shown in figure \ref{fig-HD2952}
overplotted on the velocities, whose error bars include the median value of the
derived stellar jitter of $s_{\rm slit}= 13.3$ m s$^{-1}$ and $s_{\rm fiber}= 6.5$ m s$^{-1}$
for HIDES-Slit and HIDES-Fiber mode, respectively. The rms scatter of the residuals
to the Keplerian fit was 12.4 m s$^{-1}$ and we found no significant periodicity in the
residuals at this stage, although some possible long-term trend is apparently
seen. Adopting a stellar mass of 2.54 $M_{\odot}$, we obtain
a minimum mass $m_2\sin i=1.6~M_{\rm J}$ and a semimajor axis $a=1.2$ AU
for the orbiting planet.

\begin{figure}
  \begin{center}
    \FigureFile(85mm,80mm){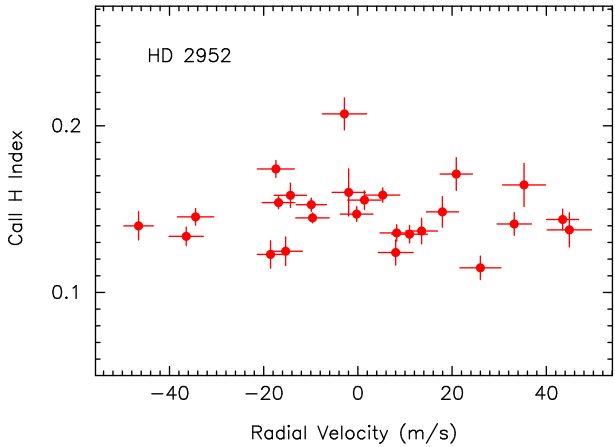}
  \end{center}
\caption{CaII H index against radial velocity
for HD 2952.}
\label{fig-HD2952_CaII}
\end{figure}

We did not find any significant variations in Ca II H index
(figure \ref{fig-HD2952_CaII}) and spectral line profiles correlated with
the orbital period (table \ref{tbl-bisector}).

\subsection{HD 120084 (HR 5184, HIP 66903)}\label{HD120084}

We obtained a total of 33 radial velocity data of HD 120084 between
2003 March and 2012 December using HIDES-Slit mode with
S/N$=$75--230 pix$^{-1}$ at 5500${\rm \AA}$ by an exposure time 900--1800 sec.

\begin{figure}
  \begin{center}
    \FigureFile(85mm,80mm){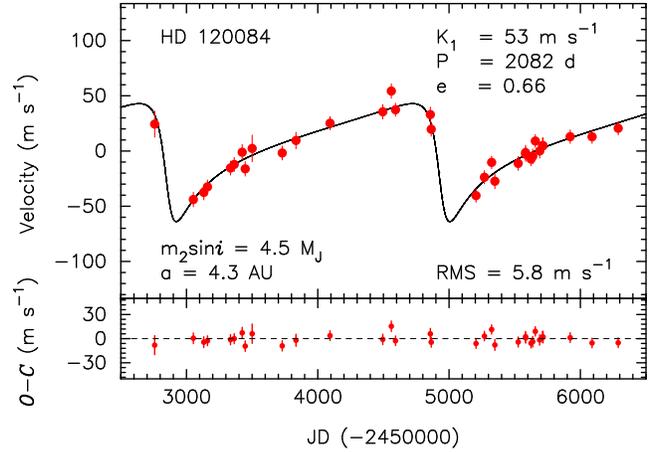}
  \end{center}
\caption{{\it Top}: Radial velocities of HD 120084 observed at OAO.
HIDES-Slit data are shown in filled red circles.
The Keplerian orbit is shown by the solid line.
The error bar for each point includes the stellar jitter.
{\it Bottom}: Residuals to the orbital fit.
The rms to the fit is 5.8 m s$^{-1}$.}
\label{fig-HD120084}
\end{figure}

The observed radial velocities are shown in figure \ref{fig-HD120084}
and are listed in table \ref{tbl-HD120084} together with their
estimated uncertainties. Lomb-Scargle periodogram of the data exhibits
a peak at around a period of 1840 days. The $FAP$ of peak is 0.06,
which is not significant because of the long period, limited phase coverage,
and deviation from sinusoidal curve for the radial velocity variations.

\begin{figure}
  \begin{center}
    \FigureFile(85mm,80mm){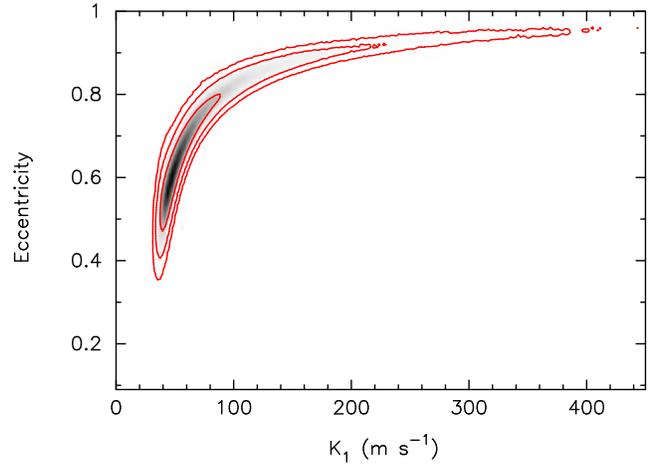}
  \end{center}
\caption{Two dimensional PDF for eccentricity $e$ and velocity semiamplitude
$K_1$ for HD 120084. The red solid lines represent contours for 68.3\%, 95.4\%,
and 99.73\% confidence from inside to out.}
\label{fig-ekHD120084}
\end{figure}

The single Keplerian model for the data  and uncertainties
for each orbital parameter were derived by MCMC method.
The median values and the 68.3\% credible regions for
orbital parameters are $P=2082^{+24}_{-35}$ d, $K_1=53^{+33}_{-11}$ m s$^{-1}$,
and $e=0.66^{+0.14}_{-0.10}$.
The resulting model is shown in figure \ref{fig-HD120084}
overplotted on the velocities, whose error bars include stellar jitter
of 5.0 m s$^{-1}$.
The rms scatter of the residuals to the Keplerian fit was
5.8 m s$^{-1}$ and we found no significant periodicity in the
residuals.
Adopting a stellar mass of 2.39 $M_{\odot}$, we obtain
$m_2\sin i=4.5~M_{\rm J}$ and $a=4.3$ AU for the companion.
Figure \ref{fig-ekHD120084} shows 2-dimensional PDF between $K_1$ and $e$.
As seen in the figure, the parameters are not well constrained yet
because the phase of velocity minimum was not covered by our observations.
However, the companion minimum-mass is below 13 $M_{\rm J}$ with 99\%
confidence, then the companion still falls into the planetary regime.

We did not analyze Ca II H variations for the star because of
the low S/N ratio for the core of the observed spectra.
We did not find any significant variations in spectral
line profiles correlated with the orbital period
(table \ref{tbl-bisector}).

\subsection{$\omega$ Serpentis (HR 5888, HD 141680, HIP 77578)}\label{oSer}

Probable periodic radial velocity variations of $\omega$ Ser were first
reported in \citet{sato:2005} at an early time of the Okayama
Planet Search Program.
A total of 123 radial velocity data of $\omega$ Ser were collected between
2001 February and 2013 January with HIDES-Slit and HIDES-Fiber mode.
We obtained S/N$=$80--340 pix$^{-1}$ at 5500${\rm \AA}$ with an exposure
time 480--1800 sec using HIDES-Slit mode and S/N$=$150--390 pix$^{-1}$ at
5500${\rm \AA}$ with an exposure time 300--900 sec using HIDES-Fiber mode.

\begin{figure}
  \begin{center}
    \FigureFile(85mm,80mm){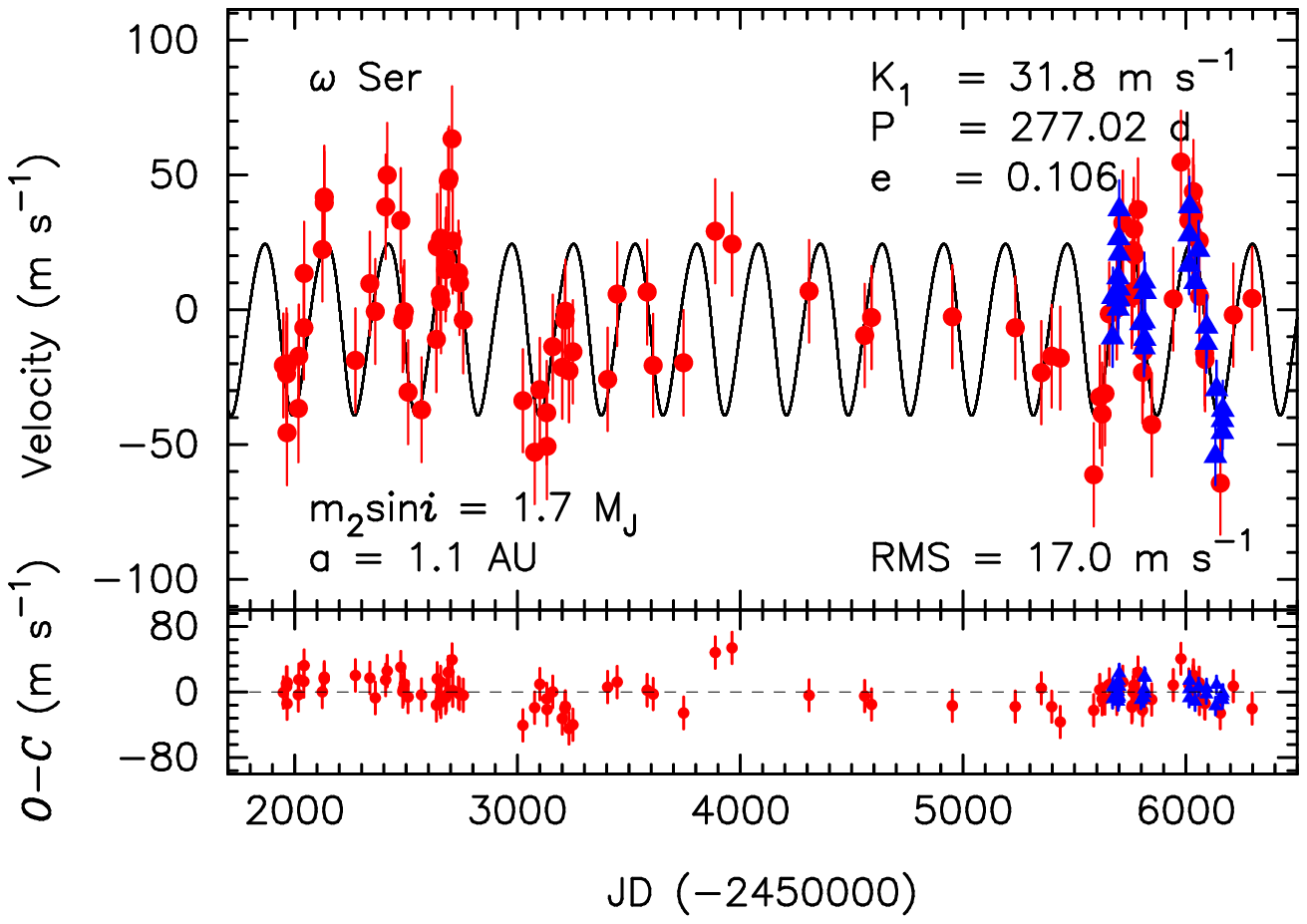}
  \end{center}
\caption{{\it Top}: Radial velocities of $\omega$ Ser observed at OAO.
HIDES-Slit data are shown in filled red circles, and
HIDES-Fiber data are filled blue triangles.
The Keplerian orbit is shown by the solid line.
The error bar for each point includes the stellar jitter.
{\it Bottom}: Residuals to the orbital fit.
The rms to the fit is 17.0 m s$^{-1}$.}
\label{fig-HD141680}
\end{figure}

The observed radial velocities are shown in figure \ref{fig-HD141680}
and are listed in table \ref{tbl-HD141680} together with their
estimated uncertainties. Lomb-Scargle periodogram of the data exhibits
a dominant peak at a period of 276 days with a $FAP<1\times10^{-5}$.

The MCMC analysis for a single Keplerian model yielded the orbital
parameters of $P=277.02^{+0.52}_{-0.51}$ d,
$K_1=31.8^{+2.3}_{-2.3}$ m s$^{-1}$, and $e=0.106^{+0.079}_{-0.069}$,
which are median and 68.3\% credible regions for PDF.
The resulting Keplerian model is shown in figure \ref{fig-HD141680}
overplotted on the velocities with error bars including the median
value of the derived stellar jitter of $s_{\rm slit}= 18.7$ m s$^{-1}$
and $s_{\rm fiber}= 10.4$ m s$^{-1}$
for HIDES-Slit and HIDES-Fiber mode, respectively.
The rms scatter of the residuals to the Keplerian fit was
17.0 m s$^{-1}$ and we found no significant periodicity in the
residuals at this stage. Adopting a stellar mass of 2.17 $M_{\odot}$, we obtain
$m_2\sin i=1.7~M_{\rm J}$ and $a=1.1$ AU for the orbiting planet.

\begin{figure}
  \begin{center}
    \FigureFile(85mm,80mm){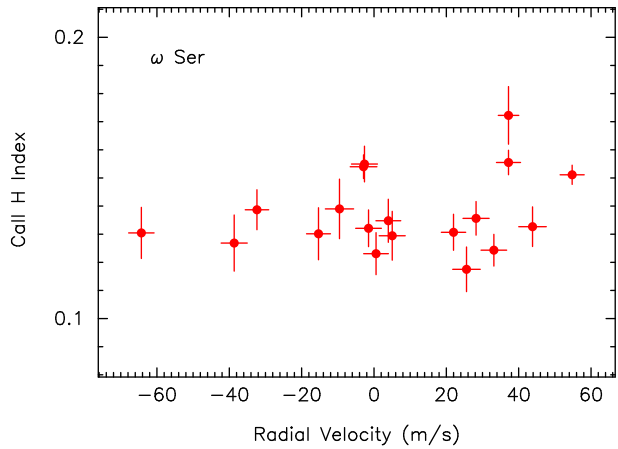}
  \end{center}
\caption{CaII H index against radial velocity
for $\omega$ Ser.}
\label{fig-HD141680_CaII}
\end{figure}

We did not find any significant variations in Ca II H index
(figure \ref{fig-HD141680_CaII}) and spectral line profiles correlated with
the orbital period (table \ref{tbl-bisector}).

\section{Discussion}\label{discussion}

\subsection{Stellar Activity}
\begin{figure}
  \begin{center}
    \FigureFile(85mm,80mm){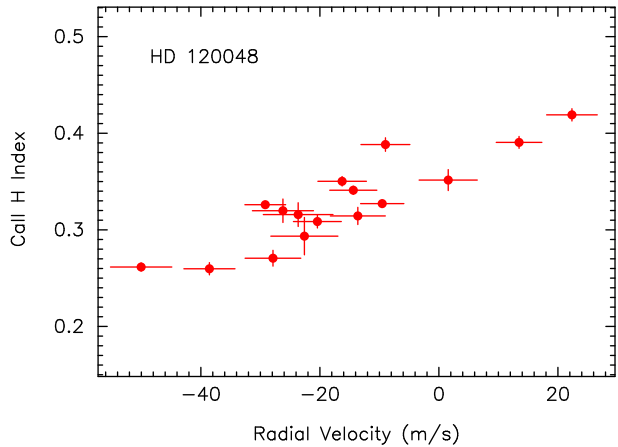}
  \end{center}
\caption{CaII H index against radial velocity
for HD 120048.}
\label{fig-HD120048_CaII}
\end{figure}

We found no variations in $S_H$ values correlated with radial velocity
variations for HD 2952, HD 120084, and $\omega$ Ser suggesting that the
observed radial velocity variations of these stars are not caused by
stellar activity. In comparison, we also examined the variations of
$S_H$ values for a chromospherically active G-type giant HD 120048,
which shows significant core reversal in the Ca II HK lines
(see figure \ref{fig-CaH}).
Figure \ref{fig-HD120048_CaII} shows the variations in $S_H$ values
for the star against radial velocity variations. It clearly shows
that the radial velocity variations are correlated with
activity level, indicating that the index can be used
to investigate a cause of radial velocity variations.

We also performed spectral line shape analysis for HD 120048
in the same manner as those for other three stars. As presented
in table \ref{tbl-bisector}, no significant variations are detected
for the star along with the other three stars. It suggests that
spectral line shape variations caused by rotational modulation
are too small to be detected by the current method or the observed
radial velocity variations in HD 120048 originate from atmospheric
motion (expansion or contraction) that are not necessarily accompanied
by significant line profile variations. The results demonstrate
the importance of monitoring not only line profile variability
but also chromospheric activity in order to investigate causes
of radial velocity variations of GK giants. Further detailed
study on the variability of HD 120048 is beyond the scope
of this paper, which will be done in a forthcoming paper.

\subsection{Detectability of Less Massive Planets around GK Giants}
HD 2952 b ($m_2\sin i=1.6~M_{\rm J}$, $a=1.2$ AU) and $\omega$ Ser b
($m_2\sin i=1.7~M_{\rm J}$, $a=1.1$ AU) are the least massive planets
ever discovered around intermediate-mass (1.5--5$M_{\odot}$) giants
together with HD 100655 b ($m_2\sin i=1.7~M_{\rm J}$; \cite{omiya:2012})
and $o$ CrB b ($m_2\sin i=1.5~M_{\rm J}$; \cite{sato:2012}).
Less massive planets are also found around low-mass ($<$1.5$M_{\odot}$)
K giants such as BD+48 738 b ($m_2\sin i=0.91~M_{\rm J}$, $a=1.0$ AU;
\cite{gettel:2012}).
It is normally more difficult to detect planets with $\lesssim2M_{\rm J}$
around such giants because of the relatively larger stellar jitter. However,
these discoveries demonstrate that it is still possible to detect such
less massive planets even around GK giants by high cadence observations.

\begin{figure}
  \begin{center}
    \FigureFile(85mm,80mm){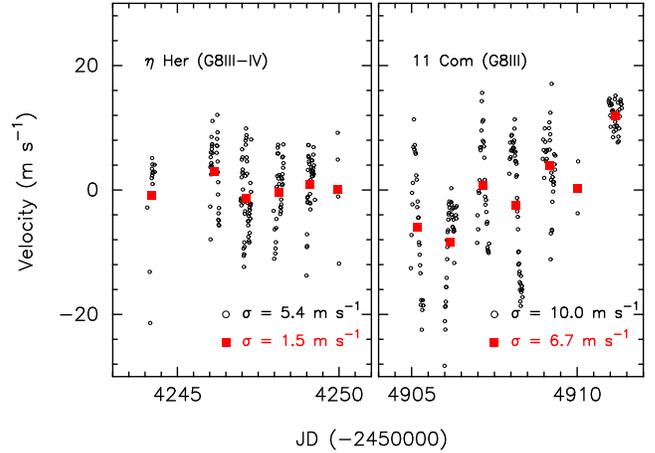}
  \end{center}
\caption{Short-term radial velocity variations for
G giants, $\eta$ Her (left) and 11 Com (right). The
radial velocity data are from \citet{ando:2010}.
Open circles are raw data and red squares are those
binned over one night. An increasing velocity trend
seen in 11 Com is caused by orbital motion (\cite{liu:2008}),
the RMS scatter of the binned data to which is 4.5 m s$^{-1}$.}
\label{fig-jitter}
\end{figure}

How low mass planets are detectable around GK giants?
\citet{ando:2010} conducted asteroseismic observations for some of
the G giants in our sample and detected solar-like oscillations in them.
The results show that the stellar jitter of the giants are primarily dominated
by solar-like oscillations with periods of 3--10 hours and integrated velocity
amplitudes of 10--20 m s$^{-1}$, which are consistent with jitters in longer
timescale (e.g., \cite{sato:2005}). In the case of solar-type dwarfs,
the p-mode oscillations have much shorter periods ($\sim$ 10 minutes) and
smaller velocity amplitudes ($\sim1$ m s$^{-1}$), which can be canceled
out down to  $\sim$20 cm s$^{-1}$ by integration for $\sim15$ minutes
(e.g., \cite{mayor:2008}). Figure \ref{fig-jitter} demonstrates the benefit
by applying the similar strategy to the radial velocity data presented
in \citet{ando:2010}.
As seen in the figure, the scatters are reduced by a factor of 2--4
by binning the data over one night ($\sim10$ hr), which makes
the detection limit of planets down to below 1 $M_{\rm J}$
at 1 AU and even super-Earth class planets in short period orbits
around stars like $\eta$ Her (G8III-IV). Although it takes time to apply this
strategy to all of our targets and the jitters may have various time
scales in variations, the results show that we can potentially
reduce the detection limit of planets around giants if we mitigate
the effect of stellar jitters by proper observational strategy.

\subsection{An Eccentric Planet: HD 120084 b}
\begin{figure}
  \begin{center}
    \FigureFile(85mm,80mm){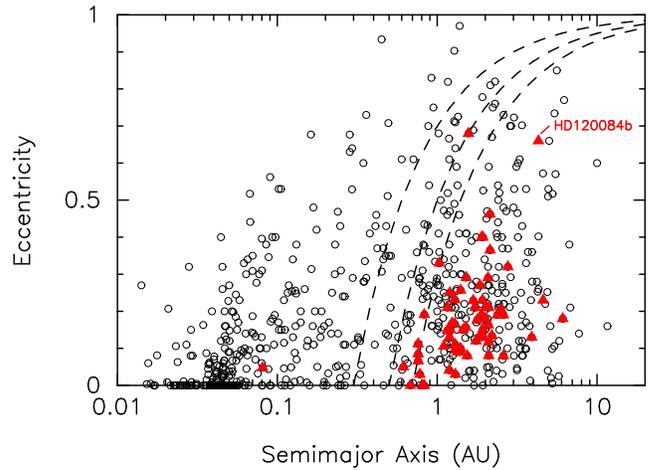}
  \end{center}
\caption{Eccentricity distribution of exoplanets
detected by radial velocity methods against semimajor
axis. The data are from http://exoplanets.eu. Red
triangles represent planets around evolved
intermediate-mass ($\ge1.5M_{\rm \odot}$) stars.
Dashed lines express the periastron distance ($q=a(1-e)$)
of 0.3, 0.5, 0.7 AU, respectively, from the left.
The planet with $e=0.68$ and $a=1.57$ AU is HD 102272c,
which is in a double planet system (\cite{niedzielski:2009a}).}
\label{fig-ecc}
\end{figure}

HD 120084 b ($m_2\sin i=4.5~M_{\rm J}$, $a=4.3$ AU) is
a long-period eccentric planet. The eccentricity exceeds
0.4 with the 99\% confidence, while almost all the planets
discovered around evolved intermediate-mass stars
have eccentricity below 0.4 (figure \ref{fig-ecc}).
Several scenarios have been proposed for the origin of eccentric planets
including planet-planet scattering (e.g., \cite{marzari:2002};
\cite{juric:2008}; \cite{ford:2008}) and secular perturbations
by an outer body (e.g., \cite{holman:1997}; \cite{mazeh:1997};
\cite{takeda:2005}), in both of which we can expect to find a distant
companion. In order to explore such a possible distant companion
around HD 120084, we performed MCMC analysis for a single
Keplerian model with a linear velocity trend $\dot{\gamma}$.
As a result, we constrained the velocity trend to be
$|\dot{\gamma}|<2$ m s$^{-1}$ yr$^{-1}$ with 99\% confidence,
giving an order-of-magnitude relation (e.g., \cite{winn:2009})
\begin{eqnarray}\label{HD5608c}
\frac{m_{\rm c} \sin i_{\rm c}}{a_{\rm c}^2} &\sim& \frac{\dot{\gamma}}{G} <
0.01~M_{\rm J}~{\rm AU^{-2}}
\end{eqnarray}
where $m_c$ is the companion mass, $i_c$ is the orbital inclination, and
$a_c$ is the orbital radius. The result could exclude existence of
a brown-dwarf companion ($\ge13~M_{\rm J}$) within $\sim$36 AU
and a stellar one ($\ge80~M_{\rm J}$) within $\sim$90 AU.

It may also be possible to assume that (unobserved) another body was
scattered into inner orbit and then engulfed by the central star.
Recently \citet{adamow:2012} reported a possible example for this
case; BD+48 740, a lithium rich giant with a possible planetary companion
in an eccentric and long-period orbit. It has been proposed for years that
the planet engulfment scenario is an origin of existence of lithium
overabundant giants (e.g., \cite{siess:1999}), and Adamow et al. suggested
that BD+48 740 might be the case.
We determined lithium abundance $A$(Li)\footnote{$A({\rm Li})=\log n_{\rm Li}/n_{\rm H}+12$}
for HD 120084 using the spectral
synthesis method with the IDL/Fortran SIU software package developed by
Reetz (1993, private communication). The synthetic spectrum is presented
by a single Gaussian profile, convolved with the broadening function of
stellar rotation, macroturbulence and instrumental profile. The lithium
abundance is obtained by fitting the synthetic spectrum for
$^7{\rm Li}~6708{\rm \AA}$ to the observed one with the contribution of
the line of FeI 6707.467${\rm \AA}$ taken into account.
We then obtained $A$(Li)$\le0.37$ for HD 120084,
which does not exhibit overabundance of lithium compared with other
giants in our sample. Details of the procedure as well as lithium abundances
for other targets of our planet-search survey can be found in Liu et al.
(in preparation).

\section{Summary}\label{summary}

Here we reported three new planetary systems around evolved
intermediate-mass stars from precise radial velocity measurements
at OAO. These add to the diversity of planets around evolved
intermediate-mass stars in terms of less massive planets than
$2~M_{\rm J}$ (HD 2952 b, $\omega$ Ser b) and a highly eccentric planet
with $e=0.66$ (HD 120084 b).

The Okayama Planet Search Program has been continuously
monitoring radial velocity of about 300 GK giants for 12 years,
which corresponds to orbital semimajor axis of 6.5 AU around
2$M_{\odot}$ stars.
Furthermore, since 2010, the high-efficiency fiber-link system for
HIDES has been available. The improvement in efficiency
allows us to observe each star more frequently, which helps to
push down the detection limit of planets. The continuous observations
combined with the frequent sampling will further uncover both
long-period and less-massive planets around evolved intermediate-mass
stars.
\\

This research is based on data collected at Okayama Astrophysical
Observatory (OAO), which is operated by National Astronomical Observatory
of Japan (NAOJ). We are grateful to all the staff members of OAO for
their support during the observations. We thank students of Tokyo
Institute of Technology and Kobe University for their kind help for
the observations. BS was partly supported by MEXT's program
"Promotion of Environmental Improvement for Independence of Young
Researchers" under the Special Coordination Funds for Promoting
Science and Technology and by Grant-in-Aid for Young Scientists (B)
20740101 from the Japan Society for the Promotion of Science (JSPS).
BS is supported by Grant-In-Aid for Scientific Research
(C) 23540263 from JSPS and HI is supported by Grant-In-Aid for
Scientific Research (A) 23244038 from JSPS.
YJL is supported by the National Natural Science Foundation of
China under grants 11173031.
This research has made use of the SIMBAD database, operated at
CDS, Strasbourg, France.


\newpage

\onecolumn
\begin{table}[h]
\caption{Stellar parameters}\label{tbl-stars}
\begin{center}
\begin{tabular}{cccccccc}\hline\hline
Parameter      & HD 2952 & HD 120084 & $\omega$ Ser\\
\hline			   			   
Sp. Type         & K0 III        & G7 III: & G8 III\\
$\pi$ (mas)     & 8.68$\pm$0.71 & 10.24$\pm$0.49 & 12.40$\pm$0.73\\
$V$                  & 5.93   & 5.91 & 5.21\\
$B-V$              & 1.037  & 1.000 & 1.019\\ 
$A_{V}$          & 0.18     & 0.00 & 0.06\\
$M_{V}$         & $+$0.44    & $+$0.96 &  $+$0.49\\
$B.C.$            & $-$0.32   & $-$0.30 & $-$0.35\\
$T_{\rm eff}$ (K) & 4844$\pm$20  & 4892$\pm$22 & 4770$\pm$10\\ 
$\log g$ (cgs)   & 2.67$\pm$0.07 & 2.71$\pm$0.08 & 2.32$\pm$0.04\\
$v_t$ (km s$^{-1}$)  & 1.32$\pm$0.08 & 1.31$\pm$0.10 & 1.34$\pm$0.04\\ 
$[$Fe/H$]$  (dex)   & $+$0.00$\pm$0.04 & $+$0.09$\pm$0.05 & $-$0.24$\pm$0.02\\
$L$ ($L_{\odot}$) & 70.8  & 43.7 & 70.8\\
$R$ ($R_{\odot}$) & 12.02 (10.96--13.18) & 9.12 (8.51--9.77) & 12.30 (11.48--13.18)\\
$M$ ($M_{\odot}$) & 2.54 (2.45--2.66) & 2.39 (2.09--2.45) & 2.17 (1.88--2.38)\\
$v\sin i$ (km s$^{-1}$) & 1.92 & 2.44 & 1.89 \\
$\sigma_{\rm HIP}$ (mag) & 0.007 & 0.007& 0.005\\
\hline
\end{tabular}
\end{center}
Note -- The uncertainties of [Fe/H], $T_{\rm eff}$, $\log g$,
and $v_{\rm t}$, are internal statistical errors (for a given dataset
of Fe~{\sc i} and Fe~{\sc ii} line equivalent widths) evaluated
by the procedure described in subsection 5.2 of \citet{takeda:2002}.
Since these parameter values are sensitive to slight changes in the
equivalent widths as well as to the adopted set of lines (\cite{takeda:2008})
realistic ambiguities may be by a factor of $\sim$ 2--3 larger
than these estimates from a conservative point of view
(e.g., 50--100 $K$ in $T_{\rm eff}$, 0.1--0.2~dex in $\log g$).
Values in the parenthesis for stellar radius and mass correspond to
the range of the values assuming the realistic uncertainties in
$\Delta\log L$ corresponding to parallax errors in the Hipparcos
catalog, $\Delta\log T_{\rm eff}$ of $\pm0.01$ dex ($\sim\pm100$~K),
and $\Delta{\rm [Fe/H]}$ of $\pm0.1$ dex.
The resulting mass value may also appreciably depend on the chosen
set of theoretical evolutionary tracks (e.g., the systematic
difference as large as $\sim 0.5M_{\odot}$ for the case of
metal-poor tracks between \citet{lejeune:2001} and
\citet{girardi:2000}).
\end{table}

\begin{table}
  \caption{Parameter Priors for MCMC Orbital Analysis}\label{tbl-priors}
  \begin{center}
    \begin{tabular}{lcrr}
  \hline\hline
  Parameter & Prior & Minimum & Maximum\\
  \hline
  $P$ (days) & Jeffreys & 1 & 10000\\
  $K_1$ (m s$^{-1}$) & Modified Jeffreys & 0 ($K_a=5$) & 1000\\
  $e$ & Uniform & 0 & 1\\
  $\omega$ & Uniform & 0 & 2$\pi$ \\
  $\chi$ & Uniform & 0 & 1 \\
  $V$ (m s$^{-1}$) & Uniform & $-$100 & 100 \\
  $s$ (m s$^{-1}$) & Modified Jeffreys & 0 ($s_a=1$) & 100\\
  \hline
    \end{tabular}
  \end{center}
\end{table}

\begin{longtable}{cccc}
  \caption{Radial Velocities of HD 2952}\label{tbl-HD2952}
  \hline\hline
  JD & Radial Velocity & Uncertainty & Obs. Mode\\
  ($-$2450000) & (m s$^{-1}$) & (m s$^{-1}$)\\
  \hline
  \endhead
3024.97290 & $-$22.7 & 3.5 & Slit\\
3025.97090 & $-$42.4 & 3.7 & Slit\\
3246.14274 & $-$17.2 & 4.1 & Slit\\
3332.97862 & $-$35.4 & 3.7 & Slit\\
3406.94360 & 0.5 & 4.0 & Slit\\
3607.23421 & $-$5.7 & 3.9 & Slit\\
3694.19166 & 8.5 & 4.2 & Slit\\
3744.01898 & 19.5 & 3.3 & Slit\\
4018.30303 & $-$3.9 & 4.3 & Slit\\
4128.91328 & 46.5 & 8.8 & Slit\\
4147.92413 & 27.8 & 11.0 & Slit\\
4339.27515 & $-$13.3 & 3.4 & Slit\\
4757.00848 & 17.9 & 3.4 & Slit\\
5135.22351 & $-$2.0 & 3.7 & Slit\\
5166.06192 & $-$34.5 & 3.9 & Slit\\
5185.98912 & $-$36.4 & 3.6 & Slit\\
5203.98979 & $-$46.6 & 3.1 & Slit\\
5233.95877 & $-$18.5 & 2.8 & Slit\\
5351.27989 & 46.7 & 3.8 & Slit\\
5398.25033 & 43.5 & 3.4 & Slit\\
5437.15263 & 8.2 & 3.5 & Slit\\
5468.21069 & $-$17.4 & 3.9 & Slit\\
5499.00815 & $-$14.4 & 5.1 & Slit\\
5524.96261 & $-$16.9 & 3.5 & Slit\\
5545.07666 & $-$11.0 & 4.7 & Slit\\
5579.98277 & 11.4 & 4.5 & Slit\\
5581.95150 & 20.9 & 3.5 & Slit\\
5582.98312 & 6.1 & 3.4 & Slit\\
5587.99681 & 14.8 & 4.0 & Slit\\
5610.92984 & 10.7 & 4.6 & Slit\\
5624.91445 & 8.6 & 4.0 & Slit\\
5758.19046 & $-$2.0 & 3.5 & Slit\\
5759.20471 & $-$15.3 & 3.6 & Slit\\
5770.26968 & $-$0.3 & 3.5 & Slit\\
5785.28326 & 5.3 & 3.7 & Slit\\
5789.19561 & $-$35.9 & 4.3 & Slit\\
5791.31690 & $-$2.9 & 4.7 & Slit\\
5810.31005 & $-$9.6 & 3.6 & Slit\\
5811.25892 & $-$9.9 & 3.2 & Slit\\
5847.04083 & 1.6 & 4.1 & Slit\\
5879.06072 & $-$20.9 & 4.4 & Slit\\
5918.88801 & $-$4.5 & 4.9 & Slit\\
5922.97437 & 8.0 & 3.7 & Slit\\
5938.92083 & 8.6 & 3.3 & Slit\\
5943.88388 & 13.5 & 3.4 & Slit\\
6088.28662 & $-$19.3 & 3.7 & Slit\\
6156.30600 & $-$14.3 & 3.4 & Slit\\
6157.21819 & 1.4 & 3.6 & Slit\\
6213.20713 & 11.0 & 3.8 & Slit\\
6235.00453 & 22.7 & 4.3 & Slit\\
6238.13304 & 35.3 & 4.6 & Slit\\
6247.98096 & 44.9 & 4.7 & Slit\\
6252.07358 & 26.0 & 4.4 & Slit\\
6283.88204 & 33.2 & 3.6 & Slit\\
  \hline
 & & &\\
6134.27677 & $-$9.1 & 2.5 & Fiber\\
6135.19472 & 1.3 & 2.6 & Fiber\\
6136.16962 & $-$17.0 & 2.4 & Fiber\\
6137.32056 & $-$18.4 & 2.3 & Fiber\\
6138.27673 & $-$10.3 & 2.4 & Fiber\\
6165.14597 & $-$4.2 & 2.8 & Fiber\\
6219.12083 & 13.4 & 4.0 & Fiber\\
6239.98610 & 27.2 & 3.8 & Fiber\\
6244.03652 & 16.9 & 3.7 & Fiber\\
  \hline
\end{longtable}

\begin{longtable}{cccc}
  \caption{Radial Velocities of HD 120084}\label{tbl-HD120084}
  \hline\hline
  JD & Radial Velocity & Uncertainty & Obs. Mode\\
  ($-$2450000) & (m s$^{-1}$) & (m s$^{-1}$)\\
  \hline
  \endhead
2758.08834 & 24.4 & 10.8 & Slit\\
3052.17823 & $-$43.9 & 4.3 & Slit\\
3131.98703 & $-$37.5 & 4.4 & Slit\\
3160.03439 & $-$32.5 & 3.6 & Slit\\
3335.31086 & $-$15.4 & 3.9 & Slit\\
3363.38688 & $-$12.2 & 3.8 & Slit\\
3424.14553 & $-$1.1 & 4.8 & Slit\\
3447.22989 & $-$16.1 & 4.2 & Slit\\
3500.09881 & 2.3 & 11.0 & Slit\\
3729.35861 & $-$1.8 & 3.8 & Slit\\
3834.18777 & 9.5 & 5.6 & Slit\\
4093.33261 & 25.2 & 3.3 & Slit\\
4495.32067 & 35.5 & 4.1 & Slit\\
4560.13449 & 54.3 & 4.2 & Slit\\
4592.13620 & 37.4 & 3.2 & Slit\\
4857.17673 & 33.1 & 4.5 & Slit\\
4864.29286 & 19.7 & 3.8 & Slit\\
5205.25700 & $-$40.4 & 3.4 & Slit\\
5269.20359 & $-$23.8 & 3.1 & Slit\\
5324.04791 & $-$10.3 & 2.8 & Slit\\
5349.06372 & $-$27.3 & 4.6 & Slit\\
5526.36838 & $-$11.2 & 3.8 & Slit\\
5582.23410 & $-$1.4 & 4.3 & Slit\\
5583.34207 & $-$2.9 & 4.0 & Slit\\
5624.30459 & $-$7.3 & 3.2 & Slit\\
5637.10714 & $-$4.8 & 4.3 & Slit\\
5657.10287 & 9.1 & 3.0 & Slit\\
5690.13478 & 0.0 & 4.0 & Slit\\
5712.07387 & 5.0 & 4.7 & Slit\\
5716.03317 & 4.8 & 5.1 & Slit\\
5922.30833 & 13.0 & 3.5 & Slit\\
6089.02875 & 12.9 & 3.3 & Slit\\
6288.33466 & 20.6 & 3.2 & Slit\\
  \hline
\end{longtable}

\begin{longtable}{cccc}
  \caption{Radial Velocities of $\omega$ Ser}\label{tbl-HD141680}
  \hline\hline
  JD & Radial Velocity & Uncertainty & Obs. Mode\\
  ($-$2450000) & (m s$^{-1}$) & (m s$^{-1}$)\\
  \hline
  \endhead
1949.32661 & $-$20.7 & 4.4 & Slit\\
1963.23330 & $-$20.0 & 8.4 & Slit\\
1963.31681 & $-$23.8 & 4.7 & Slit\\
1965.33358 & $-$45.6 & 5.1 & Slit\\
1966.30201 & $-$20.7 & 4.3 & Slit\\
2016.29162 & $-$36.6 & 7.0 & Slit\\
2018.19403 & $-$17.2 & 3.8 & Slit\\
2041.13940 & $-$6.7 & 3.9 & Slit\\
2042.11952 & 13.5 & 3.8 & Slit\\
2123.98945 & 22.3 & 4.4 & Slit\\
2132.99172 & 39.6 & 3.0 & Slit\\
2133.00942 & 41.8 & 3.2 & Slit\\
2272.35035 & $-$18.8 & 4.6 & Slit\\
2337.31840 & 9.7 & 4.4 & Slit\\
2361.31453 & $-$0.6 & 5.1 & Slit\\
2408.08042 & 38.2 & 4.9 & Slit\\
2415.09944 & 49.9 & 4.9 & Slit\\
2475.99488 & 33.1 & 4.9 & Slit\\
2485.04153 & $-$3.9 & 4.0 & Slit\\
2488.04647 & $-$2.9 & 3.3 & Slit\\
2492.01853 & $-$0.7 & 3.3 & Slit\\
2508.98656 & $-$30.6 & 4.5 & Slit\\
2568.89966 & $-$37.1 & 5.1 & Slit\\
2635.37474 & $-$11.0 & 5.6 & Slit\\
2639.38303 & 23.3 & 5.8 & Slit\\
2653.37319 & 5.7 & 4.3 & Slit\\
2654.38879 & 26.5 & 4.8 & Slit\\
2656.34630 & 3.1 & 4.2 & Slit\\
2677.35725 & 14.9 & 4.0 & Slit\\
2680.34513 & 18.8 & 4.2 & Slit\\
2689.32234 & 47.8 & 5.1 & Slit\\
2692.28082 & 48.8 & 4.1 & Slit\\
2706.26453 & 63.4 & 5.3 & Slit\\
2709.31474 & 25.5 & 5.3 & Slit\\
2735.22182 & 13.8 & 4.6 & Slit\\
2739.27107 & 9.9 & 4.5 & Slit\\
2756.25351 & $-$3.8 & 6.4 & Slit\\
3024.37398 & $-$33.7 & 3.8 & Slit\\
3077.25145 & $-$52.8 & 4.4 & Slit\\
3100.25088 & $-$29.6 & 4.1 & Slit\\
3131.13458 & $-$38.2 & 4.1 & Slit\\
3132.27689 & $-$50.6 & 5.9 & Slit\\
3158.18452 & $-$13.7 & 4.5 & Slit\\
3201.07394 & $-$21.3 & 3.8 & Slit\\
3213.02084 & $-$3.8 & 4.0 & Slit\\
3215.04862 & $-$0.6 & 3.8 & Slit\\
3231.00895 & $-$22.7 & 3.8 & Slit\\
3248.98212 & $-$15.6 & 3.8 & Slit\\
3404.34420 & $-$25.8 & 3.9 & Slit\\
3447.25348 & 5.8 & 4.3 & Slit\\
3581.98388 & 6.6 & 5.0 & Slit\\
3609.01868 & $-$20.6 & 3.8 & Slit\\
3745.37391 & $-$19.6 & 5.6 & Slit\\
3887.22571 & 29.1 & 4.5 & Slit\\
3962.95693 & 24.3 & 3.7 & Slit\\
4308.99011 & 6.9 & 3.2 & Slit\\
4558.19006 & $-$9.5 & 3.9 & Slit\\
4589.19334 & $-$2.9 & 3.7 & Slit\\
4952.20424 & $-$2.6 & 3.6 & Slit\\
5234.30043 & $-$6.7 & 3.3 & Slit\\
5351.02660 & $-$23.3 & 3.8 & Slit\\
5398.04047 & $-$17.2 & 3.4 & Slit\\
5435.96330 & $-$17.9 & 3.6 & Slit\\
5586.31423 & $-$61.2 & 3.9 & Slit\\
5613.33791 & $-$32.3 & 3.2 & Slit\\
5624.36003 & $-$38.6 & 3.6 & Slit\\
5636.34089 & $-$31.1 & 4.3 & Slit\\
5656.22321 & $-$1.5 & 3.6 & Slit\\
5690.16462 & 0.6 & 3.4 & Slit\\
5717.09514 & 32.2 & 4.8 & Slit\\
5757.06063 & 4.8 & 3.8 & Slit\\
5759.08554 & 6.7 & 3.7 & Slit\\
5763.01162 & 22.0 & 3.4 & Slit\\
5766.00806 & 29.8 & 3.7 & Slit\\
5770.05598 & 20.2 & 3.8 & Slit\\
5785.01129 & 37.2 & 2.8 & Slit\\
5787.96807 & 8.2 & 4.5 & Slit\\
5804.95034 & $-$23.2 & 2.8 & Slit\\
5809.94684 & $-$15.3 & 3.3 & Slit\\
5846.89877 & $-$42.6 & 4.3 & Slit\\
5943.37404 & 4.0 & 3.4 & Slit\\
5978.36886 & 54.8 & 3.3 & Slit\\
6013.26399 & 33.1 & 3.5 & Slit\\
6030.28218 & 28.2 & 3.6 & Slit\\
6032.20041 & 37.2 & 3.3 & Slit\\
6034.20383 & 43.8 & 3.8 & Slit\\
6035.31167 & 34.6 & 3.6 & Slit\\
6060.10402 & 25.6 & 3.8 & Slit\\
6061.09811 & 5.0 & 3.6 & Slit\\
6084.08376 & $-$16.2 & 4.2 & Slit\\
6086.18326 & $-$18.4 & 4.1 & Slit\\
6154.97320 & $-$64.3 & 3.5 & Slit\\
6212.90005 & $-$2.0 & 3.7 & Slit\\
6297.38811 & 4.2 & 3.7 & Slit\\
  \hline
 & & &\\
5671.20537 & $-$9.5 & 3.2 & Fiber\\
5672.26886 & 5.5 & 3.3 & Fiber\\
5695.15717 & 4.6 & 2.8 & Fiber\\
5695.28079 & 12.8 & 2.9 & Fiber\\
5696.06702 & 1.1 & 2.8 & Fiber\\
5696.10871 & 4.9 & 2.6 & Fiber\\
5697.08506 & 5.0 & 2.5 & Fiber\\
5698.21953 & 27.3 & 4.0 & Fiber\\
5700.15563 & 8.6 & 3.1 & Fiber\\
5701.02381 & 38.0 & 3.1 & Fiber\\
5701.12905 & 21.5 & 2.6 & Fiber\\
5797.95807 & $-$4.2 & 2.4 & Fiber\\
5811.94795 & $-$13.2 & 2.2 & Fiber\\
5812.96282 & $-$3.5 & 2.4 & Fiber\\
5813.94961 & 11.3 & 2.8 & Fiber\\
5814.93863 & $-$10.0 & 3.0 & Fiber\\
5816.97108 & 7.3 & 2.7 & Fiber\\
6014.22031 & 17.4 & 3.1 & Fiber\\
6015.17803 & 28.6 & 3.0 & Fiber\\
6016.16312 & 39.2 & 3.5 & Fiber\\
6041.26690 & 11.1 & 3.2 & Fiber\\
6057.22610 & 23.0 & 2.7 & Fiber\\
6092.06771 & $-$11.5 & 3.5 & Fiber\\
6093.16308 & $-$5.5 & 5.3 & Fiber\\
6133.01204 & $-$53.5 & 2.5 & Fiber\\
6137.96888 & $-$28.7 & 2.2 & Fiber\\
6164.00431 & $-$44.6 & 2.5 & Fiber\\
6164.99025 & $-$40.1 & 2.6 & Fiber\\
6165.99606 & $-$36.3 & 2.8 & Fiber\\
  \hline
\end{longtable}

\begin{table}
  \caption{Orbital Parameters}\label{tbl-planets}
  \begin{center}
    \begin{tabular}{lrrr}
  \hline\hline
  Parameter      & HD 2952 b & HD 120084 b & $\omega$ Ser b\\
  \hline
$P$ (days)         &  311.6$^{+1.7}_{-1.9}$ & 2082$^{+24}_{-35}$ & 277.02$^{+0.52}_{-0.51}$ \\
$K_1$ (m s$^{-1}$)  &  26.3$^{+3.8}_{-3.4}$  & 53$^{+33}_{-11}$ & 31.8$^{+2.3}_{-2.3}$\\
$e$                &  0.129$^{+0.099}_{-0.085}$ & 0.66$^{+0.14}_{-0.10}$ & 0.106$^{+0.079}_{-0.069}$\\
$\omega$ (deg)     &  64$^{+56}_{-48}$ & 117$^{+12}_{-9}$ & 132$^{+34}_{-52}$\\
$T_p$    (JD$-$2,450,000)  & 112$^{+64}_{-54}$ & 774$^{+80}_{-61}$ & 22$^{+27}_{-38}$\\
$s_{\rm slit}$ (m s$^{-1}$) & 13.3$^{+1.6}_{-1.4}$  & 5.0$^{+1.2}_{-1.0}$ & 18.7$^{+1.6}_{-1.4}$\\
$s_{\rm fiber}$ (m s$^{-1}$) & 6.5$^{+2.6}_{-1.8}$  & -- & 10.4$^{+1.8}_{-1.4}$\\
$\Delta$RV$_{\rm f-s}$ (m s$^{-1}$) & 6.9$^{+3.8}_{-3.8}$ & -- & 0.84$^{+2.9}_{-2.9}$\\
$a_1\sin i$ (10$^{-3}$AU)     & 0.74$^{+0.10}_{-0.09}$ & 7.7$^{+2.4}_{-1.1}$ & 0.804$^{+0.057}_{-0.057}$\\
$f_1(m)$ (10$^{-7}M_{\odot}$) & 0.0057$^{+0.0027}_{-0.0019}$ & 0.14$^{+0.18}_{-0.05}$ & 0.0090$^{+0.0021}_{-0.0018}$\\
$m_2\sin i$ ($M_{\rm J}$)  & 1.6  & 4.5 & 1.7\\
$a$ (AU)                  & 1.2  & 4.3  & 1.1\\
$N_{\rm slit}$              & 54   & 33  & 94\\
$N_{\rm fiber}$              & 9   & --  & 29\\
RMS (m s$^{-1}$)           & 12.4  & 5.8  & 17.0\\
  \hline
    \end{tabular}
  \end{center}
\end{table}

\begin{table}
  \caption{Bisector Quantities}\label{tbl-bisector}
  \begin{center}
    \begin{tabular}{lrrrr}
  \hline\hline
  Bisector Quantities & HD 2952 & HD 120084 & $\omega$ Ser & HD 120048\\
  \hline
Bisector Velocity Span (BVS) (m s$^{-1}$) & $-$6.0$\pm$2.9 & 1.3$\pm$3.7 & 2.1$\pm$4.2 & 2.0$\pm$5.1\\
Bisector Velocity Curvature (BVC) (m s$^{-1}$) & 1.2$\pm$1.3 & $-$0.9$\pm$2.2 & 0.6$\pm$2.1 & $-$4.8$\pm$3.0\\
Bisector Velocity Displacement (BVD) (m s$^{-1}$) & $-$69.6$\pm$5.1 & $-$70.0$\pm$6.2 & $-$102.4$\pm$8.5 & $-$68.0$\pm$8.8\\
  \hline
    \end{tabular}
  \end{center}
\end{table}

\end{document}